\documentclass{aa}
\usepackage{graphicx}
\def\approxgt{\mathrel{\hbox{\rlap{\lower.55ex \hbox {$\sim$}}
        \kern-.3em \raise.4ex \hbox{$>$}}}}
\def\approxlt{\mathrel{\hbox{\rlap{\lower.55ex \hbox {$\sim$}}
        \kern-.3em \raise.4ex \hbox{$<$}}}}

\begin{document}
   \title{The early stage of a cosmic collision? XMM-Newton unveils two obscured AGN in the galaxy pair ESO509-IG066}

   \author{Matteo Guainazzi
          \inst{1},
	  Enrico Piconcelli,
	  \inst{1}
	  Elena Jim\'enez-Bail\'on,
	  \inst{1}
	  Giorgio Matt
	  \inst{2}
          }

   \offprints{M.Guainazzi}

   \institute{$^1$XMM-Newton Science Operation Center, European Space Astronomy Center, ESA, Apartado
              50727, E-28080 Madrid, Spain \\
              \email{mguainaz@xmm.vilspa.esa.es} \\
	      $^2$Dipartimento di Fisica ``E.Amaldi'', Universit\`a ``Roma
		Tre'', Via della Vasca Navale 84, I-00146, Roma, Italy
              }

   \date{Received ; accepted }

   \abstract{We report the XMM-Newton discovery of a X-ray
	    bright AGN pair in the
	    interacting galaxy system ESO509-IG066.
	    Both galaxies host an X-ray luminous
	    ($L_{\rm X} \sim 10^{43}$~erg~s$^{-1}$) obscured nucleus
	    with column densities
	    $N_{\rm H} \simeq 7 \times 10^{22}$~cm$^{-2}$ and
	    $N_{\rm H} \simeq 5 \times 10^{21}$~cm$^{-2}$. The
	    optical morphology is only mildly disturbed, suggesting a
	    merging system in the early stage of its evolution.
            Still, the pair is probably gravitationally bound,
	    and might eventually evolve into a compact, fully gas
	    embedded systems such as NGC~6240 (\cite{komossa03}).  
   \keywords{Galaxies:interactions --
             Galaxies:Seyfert --
	     Galaxies:individual:ESO509-IG066 --
	     X-rays:galaxies
            }
            }

\authorrunning{Guainazzi et al.}

\titlerunning{X-ray from the galaxy pair ESO509-IG066}

   \maketitle
%

\section{Introduction}

About 20 {\it bona fide} Active Galactic Nuclei
(AGN) pairs are currently known
(\cite{kochanek99}).
They represent about
0.1\% of QSO optical surveys (\cite{hewett98}), although
this number is dependent on the criteria used to
distinguish between
``true pairs'' and gravitational
lenses (\cite{mortlock99}).

AGN pairs are a potentially interesting laboratory to study the
early phases of AGN activity. Gas shock and compression
caused by galaxy interactions may
lead to feeding otherwise quiescent
super-massive black holes, and to enhanced
star formation (\cite{rees84,byrd86}).
The possible role of merging and flybys to bring gas to the
nuclear region has been examined by means of
N-body simulations
(\cite{barnes92,hernquist95,taniguchi96}). Observationally,
quasars seem indeed to live in denser environment then
normal galaxies
(\cite{kauffmann04}), whereas the same
evidence for low-luminosity
AGN is still controversial
(\cite{laurikainen94,rafanelli95,derobertis98,schmitt01}).

In this paper we present the first
X-ray imaging and spectroscopic observation of the
interacting galaxy pair ESO509-IG066
(\cite{arp87}), and report
the discovery that {\it both} galaxies host
a luminous
($L_{\rm X} \sim 10^{43}$~erg~s$^{-1}$) and obscured X-ray
source.
The optical nuclei of the pair are
\begin{figure}
   \centering
{\it File Gi131.gif}
\caption{
HST/WFC2 image of the ESO509-IG066 field
in the F606W filter.
The {\it circles} represent the EPIC/XMM-Newton
sources error boxes.
              }
\label{fig4}
\end{figure}
aligned in the E-W direction at a projected
separation of $\simeq$16$\arcsec$, with spectroscopic redshifts
$z_{\rm E} = 0.033223 \pm 0.00003$ and $z_{\rm W} = 0.034307 \pm 0.00014$
(\cite{sekiguchi92}). The W source was classified as a Seyfert~2
galaxies on the basis of the [N{\sc ii}]$\lambda$6583/H$_{\alpha}$
and [O{\sc iii}]$\lambda$5007/H$_{\beta}$ ratios. In the E
source, the lack of H$_{\beta}$ detection favored a H~{\sc II} or
LINER classification (\cite{sekiguchi92}). In
the X-ray band, only a detection by the {\it Ginga}/LAC
is reported in the literature, with a 2-10~keV flux
of $\simeq 7 \times 10^{-12}$~erg~cm$^{-2}$~s$^{-1}$ (\cite{polletta96}).

At their redshift,
the apparent distance of the pair members
translates into a projected
physical separation $R \simeq 10.5$~kpc.
The line-of-sight velocity difference is
$\Delta v_{||} = 320 \pm 40$~km~s$^{-1}$.
For comparison, in the Mortlock et al. (1999) sample of
binary quasars: $\langle R \rangle = 32 \pm 9$~kpc
and $\langle \Delta v_{||} \rangle = 83 \pm 10$~km~s$^{-1}$.

In this paper: energies are quoted in the source's
frame; errors on the count rates are at the 1-$\sigma$ level; errors
on the spectral parameters are at the 90\% confidence level
for 1 interesting parameter; a flat $\Lambda$CDM cosmology
with $H_{\rm 0} = 70$~Mpc~km~s$^{-1}$ and
($\Omega_{\rm M},\Omega_{\Lambda}$) = (0.3,0.7),
(\cite{bennett03}) is assumed,
unless otherwise specified.

\section{The data}

XMM-Newton observed the sky region around ESO509-IG066 on January 24,
2004. The $\simeq$30$\arcmin$$\times$30$\arcmin$ field-of-view
EPIC cameras (MOS; Turner et al. 2001; pn, St\"uder et
al. 2001) were operating in Full Frame Mode with the {\sc Medium}
and {\sc Thin} optical rejection filter, respectively.
Data were reduced with SAS v6.0.0, using
the most updated calibration files.
Particle background was screened by applying
optimized thresholds to
the single-event, $E > 10$~keV, 10-s binned, field-of-view 
light curve:
0.5 and  3.5~$s^{-1}$ for the MOS and pn,
respectively. After screening the exposure times are 9.7
and 8.6~ks for the MOS and the pn, respectively.

\begin{figure*}
   \centering
   \includegraphics[angle=-90,width=16cm]{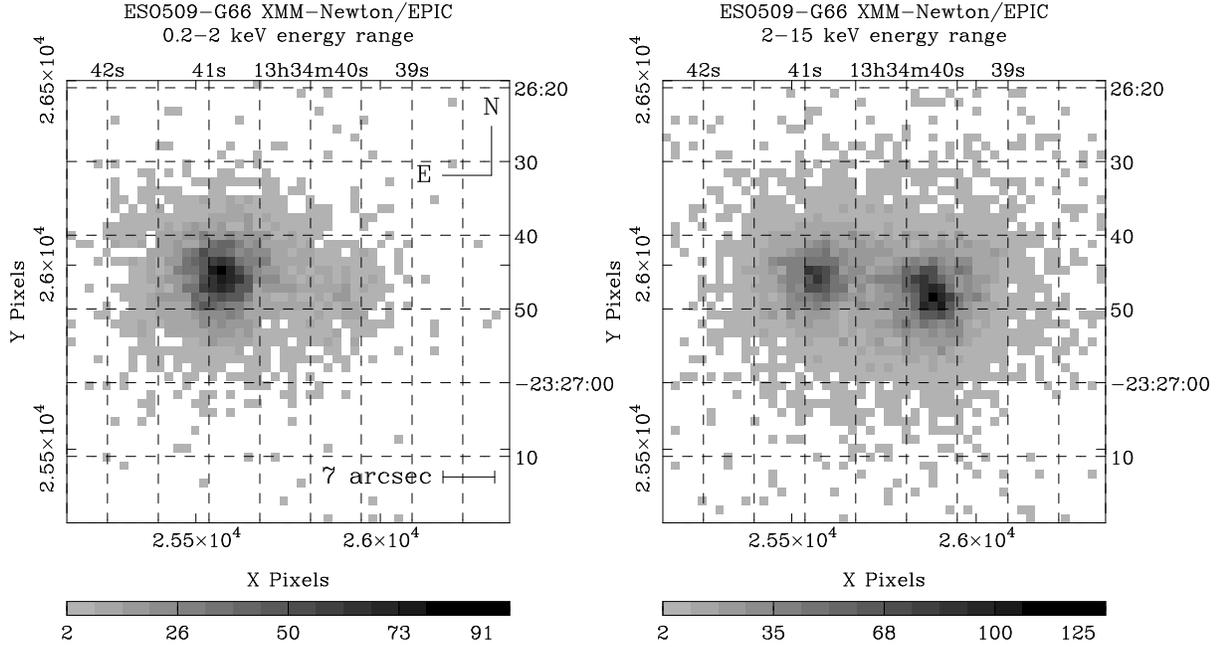}
      \caption{EPIC combined image of the innermost 1$\arcmin$
		around the ESO509-IG066 galaxy pair centroid in
		the 0.2--2~keV ({\it left panel}) and
		2--15~keV ({\it right panel}) energy bands.
		A constant level of 2 counts per pixel has been
		subtracted
              }
         \label{fig1}
\end{figure*}

In Fig.~\ref{fig1} we show the
EPIC images of the innermost 1$\arcmin$
around the ESO509-IG066 Galaxy Pair centroid in the soft
(0.2--2~keV) and hard (2--15~keV)
energy bands. In the soft band, the E source is clearly visible,
whereas the W source becomes the brightest
in the hard band.
\begin{table}
\caption{Sources detected by the EPIC camera at a signal-to-noise
level $>3$ in the innermost 1$\arcmin$ around the ESO509-IG066
galaxy pair centroid. ``E'' and ``W'' refer to the
components of the AGN pair. Count Rates (CR) refer to the combined
MOS cameras in the 0.2--15~keV band. The Error Box (EB) is
purely statistics.}
\begin{tabular}{lcccc} \hline \hline
Source & RA (J2000) & Dec (J2000) & EB & CR \\
& & & (${\arcsec}$) & (s$^{-1}$) \\ \hline
W & 13$^{\rm h}$34$^{\rm m}$39$^{\rm s}$.7 & -23$^{\circ}$26$\arcmin$48$\arcsec$ & 7 & $0.658 \pm 0.018$ \\
E & 13$^{\rm h}$34$^{\rm m}$40$^{\rm s}$.8 & -23$^{\circ}$26$\arcmin$45$\arcsec$ &  10 & $0.512 \pm 0.016$ \\ \hline \hline
\end{tabular}
\label{tab1}
\end{table}
Their positions (Tab.~\ref{tab1})
are consistent with the optical nuclei of the galaxy pair
members, once the statistical error box,
and a residual $\simeq$2$\arcsec$ systematic uncertainties
in the absolute position reconstruction are taken into
account.

For such close X-ray sources,
point spread function contamination is a potential issue in EPIC.
In order to account for this effect,
we extracted source scientific products from comparatively
small circles of 7$\arcsec$ and
8$\arcsec$ radius for Source~W and E, respectively.
We used only single and double (single to quadruple) events
in the pn (MOS).
Background scientific products
for each galaxy were generated by combining
spectra extracted from standard offset regions on the
same chip as the targets, and the appropriate fraction
(4\%) of the
spectrum of the companion. Spectra were
rebinned in order to oversample the instrumental
resolution by a factor not larger than 3, and
to ensure that
each background-subtracted spectral channel has 25 counts at
least. We have restricted the 
spectral analysis to the bands 0.5--10~keV and 
0.35--15~keV for the MOS and the pn, respectively, where the
instrument are best calibrated, and fit the spectra
simultaneously with {\sc Xspec v11.3.0}.

\begin{figure}
   \centering
   \includegraphics[angle=-90,width=8cm]{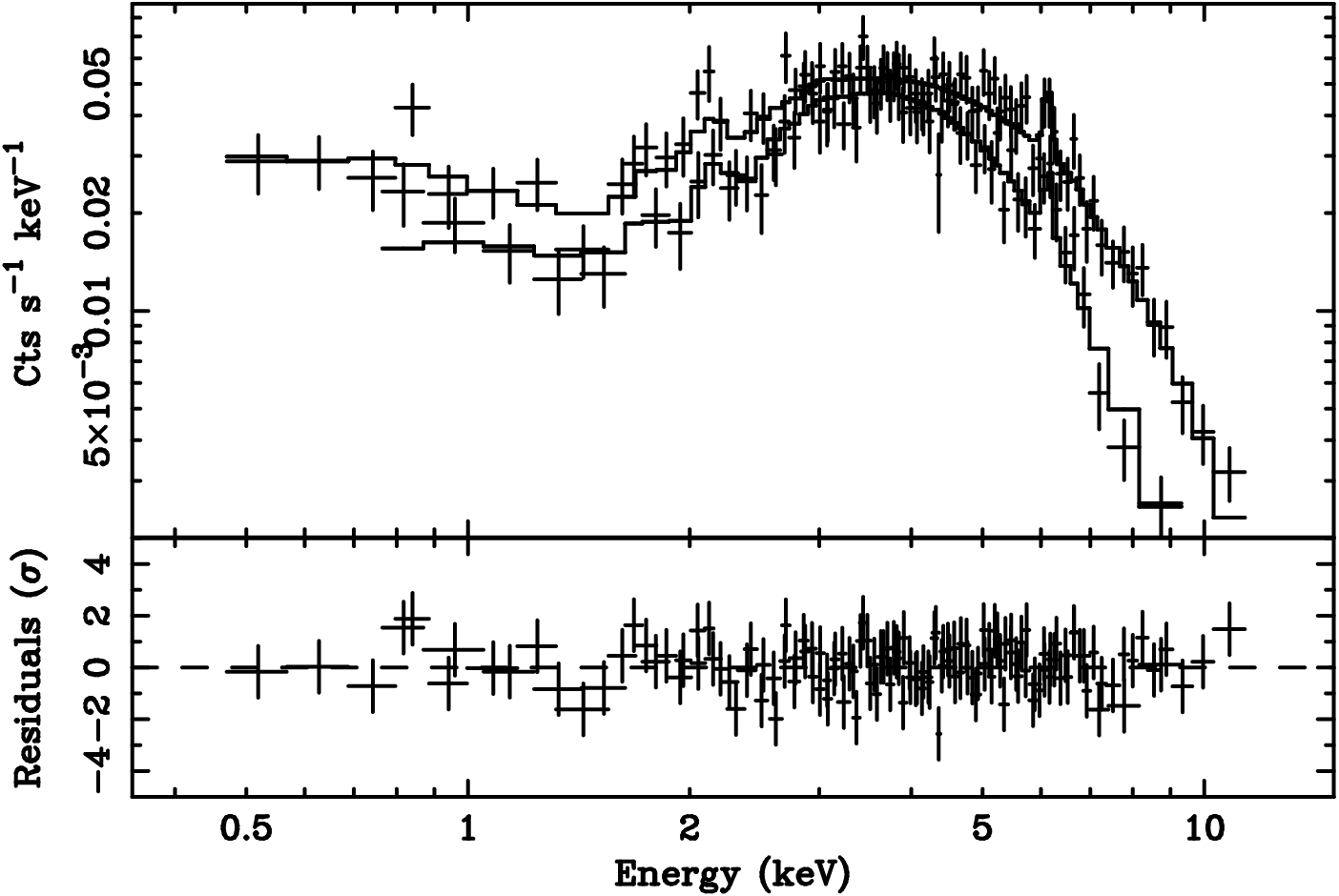}
   \includegraphics[angle=-90,width=8cm]{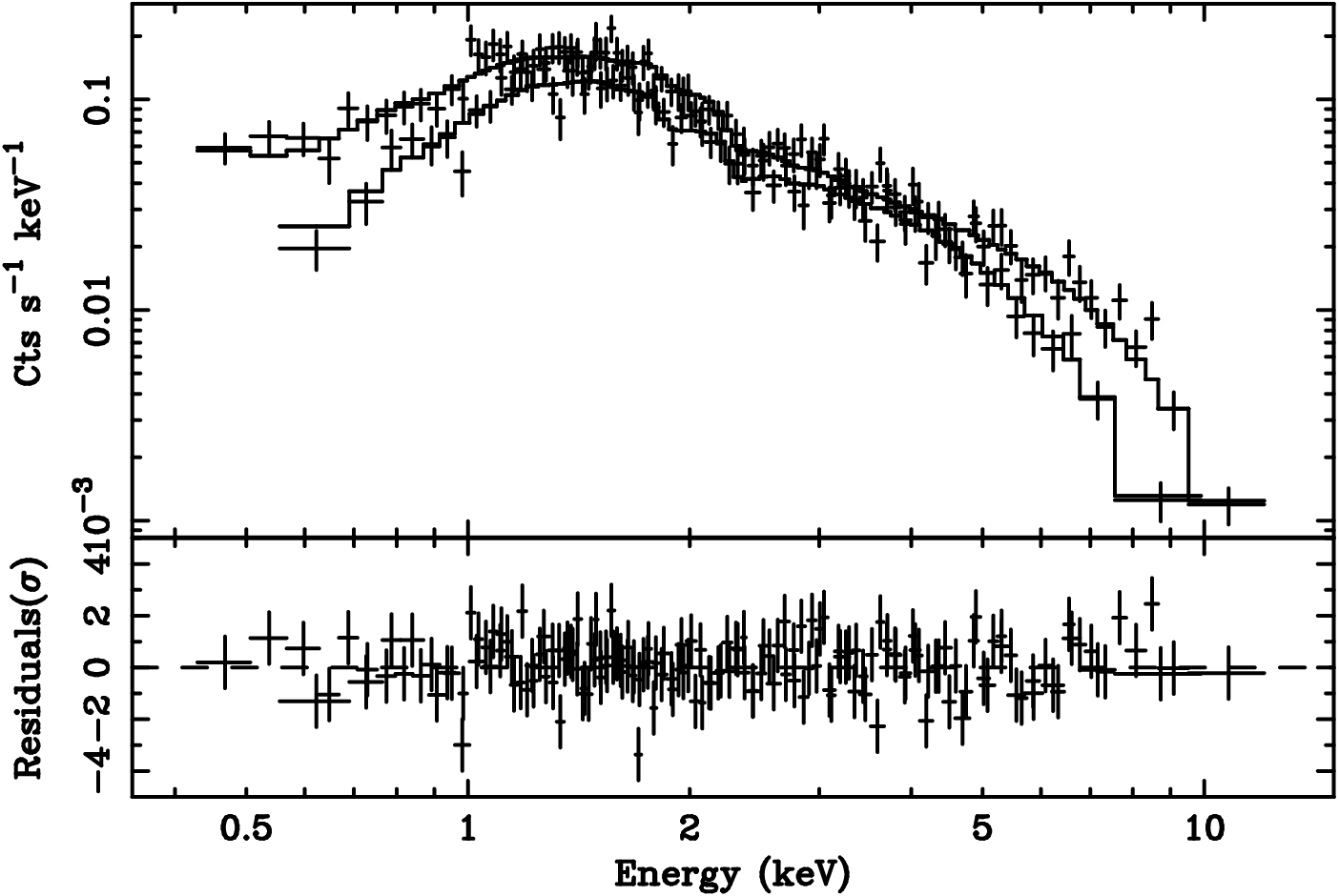}
      \caption{Spectra ({\it upper panels}) 
and residuals in units of standard deviations
({\it lower panels}) when the best-fit model is applied
to the pn and MOS spectra of source \#1 ({\it
top}) and \#2 ({\it bottom}).
              }
         \label{fig2}
\end{figure}

The spectra of both sources can be reasonably well fit with
the combination of two continua.
The bulk of the X-ray
emission is due to a photoelectrically absorbed
power-law. The column densities are
$N_{\rm H}$$\simeq$$7.1 \times 10^{22}$~cm$^{-2}$, and
$\simeq$$5.9 \times 10^{21}$~cm$^{-2}$ for
Source~W and E, respectively. Both
are significantly larger than the contribution
due to intervening gas in our Galaxy ($N_{\rm H,Gal}
= 6.5 \times 10^{20}$~cm$^{-2}$, Dickey \& Lockman 1990).
A soft excess
above the photoelectrically absorbed power-law
can be well accounted for by another power-law
modified only by Galactic absorption
In Source~W an emission line is detected at
the 99.95\% confidence level according
to the F-test (\cite{protassov02}). Its
centroid energy is consistent with K$_{\alpha}$
fluorescence from neutral or mildly ionized iron.
The Equivalent Width ($EW \simeq 120$~eV) is
typical of Compton-thin Seyfert~2 galaxies
(\cite{risaliti02}).
Tab.~\ref{tab2} lists the best-fit parameters
\begin{table*}
\caption{Best-fit parameters and results for the EPIC
spectra of the ESO509-IG066 Galaxy Pair members. $f_{\rm s}$ is the
scattering fraction ($1-f_{\rm s}$ is the absorber covering fraction in the
``leaky absorber'' scenario). $F$ is the observed flux in
the 0.5--2/2--10~keV energy bands. $L$ is the absorption-corrected
luminosity in the 0.5--10~keV energy band.}
\begin{tabular}{lcccccccc} \hline \hline
& \multicolumn{2}{c}{Absorbed power-law} & Soft excess & \multicolumn{2}{c}{K$_{\alpha}$ iron line} & \multicolumn{2}{c}{Fluxes and luminosities} \\
Source & $N_{\rm H}$ & $\Gamma$ & $f_{\rm s}$ & $E_{\rm c}$ & $EW$ &  F & $L$ & $\chi^2/\nu$ \\ 
& ($10^{22}$~cm$^{-2}$) & & (\%) & (keV) & (eV) &  ($10^{-12}$~erg~cm$^{-2}$~s$^{-1}$) & ($10^{43}$~erg~s$^{-1}$) & \\ \hline
W & $7.1 \pm^{0.6}_{0.5}$ & $1.81 \pm 0.13$ & $2.3 \pm^{0.5}_{0.6}$ & $6.35 \pm 0.05$ & $120 \pm^{60}_{50}$ & 0.12/5.2 & $3.3 \pm 0.4$ & 105.8/130 \\
E & $0.51 \pm^{0.46}_{0.10}$ & $1.64 \pm^{0.72}_{0.07}$ & $5.8 \pm^{14.5}_{5.2}$ & 6.4$^a$ & $<90$ & 0.6/3.0 & $1.43 \pm 0.08$ & 170.2/167 \\ \hline \hline
\end{tabular}

\noindent
$^a$fixed

\label{tab2}
\end{table*}
and results.
Fig.~\ref{fig2}
shows the spectra, best-fit models,
and residuals.

The 0.5--10~keV intrinsic luminosity
is $\sim 10^{43}$~erg~s$^{-1}$ for both X-ray sources, typical
for luminous Seyfert galaxies, and much
larger than observed in LINERS
(\cite{ho01,terashima02}). A substantial contribution
by shocked starburst gas,
X-ray binaries or supernovae
is unlikely. ESO509-IG066
is the X-ray brightest
AGN pair ever detected so far.
At the flux level observed by XMM-Newton,
both AGN should have been detected by the
ROSAT All Sky-Survey, with count rates
of $\simeq$5.3 and $1.4 \times 10^{-2}$~s$^{-1}$
for source E and W, respectively. However,
the closest detected RASS source is $\ge$50$\arcmin$
away from the AGN pair centroid. Only a count rate
3$\sigma$
upper limit of $7 \times 10^{-3}$~s$^{-1}$
was measured by the RASS at the AGN pair position.

\section{Discussion}

Once established that most
galaxies host a super-massive black hole
(\cite{kormendy97}), the next open issue is the mechanism that triggers
gas accretion and ultimately nuclear activity.
Galaxy encounters could be one
of these triggers (\cite{silk98, taniguchi99}). Gravitational
torques generated during the encounter can cause
gas inflow toward the nucleus (\cite{barnes96}). Although
such inflows are not strong enough to create {\it new} black
holes, they are consistent with the refueling
of ``quiescent'' black holes.

Is the pair in ESO509-IG066 a binary AGN?
On a statistical basis, binary quasars
become active at projected separations between 50 and
100~kpc, and recover (apparent) quiescence
at separations shorter than 10~kpc (\cite{mortlock99}). 
If the onset of nuclear activity is indeed related
to galaxy interaction, it should ``turn on'' during
galaxy merging, and AGN
should be found preferentially in highly disturbed
environment, with enhanced star formation. This is, however,
{\it not} the case of ESO509-IG066, whose galaxies
exhibit only mild surface brightness disturbances
(cf. Fig.~\ref{fig4}).
An analogous case is MGC~2214+3350 (\cite{munoz98}).
Morphologically mildly disturbed systems could represent
young binaries, in the earliest phase of their
encounter (\cite{kochanek99}). It is tempting
to speculate that they might represent the first
stage of an evolutionary sequence, at whose end one finds
compact,
morphologically highly disturbed systems like
NGC~6240 (\cite{tecza00}). This would require that
the AGN lifetime is comparable to
the time scale for the orbital
decay ($\sim 10^9 M^{-1}_9$~yr, where
$M_9$ is the black hole mass in units of $10^9 M_{\odot}$,
\cite{binney87}). 

Whatever its ultimate fate, the AGN pair in
ESO509-IG066 is likely to be a gravitationally
bound system.
The condition on the minimum center-of-mass frame energy
being $\le 0$ (\cite{mortlock99}) implies a lower limit on
the total mass of the system: $M_{\rm cr} \simeq (R \Delta v_{||}^2)/(2 G)
\simeq 2 \times 10^{11} M_{\odot}$.
From the HST image, we estimate a bulge V luminosity of
8.7 and $3.9 \times 10^{43}$~erg~s$^{-1}$ for the E
and W source, respectively, corresponding to
a total black hole mass $\sim$$1.2 \times 10^9 M_{\odot}$
(\cite{magorrian98}). This in turn translates into an
estimated dark matter halo mass exceeding $10^{13}$~M$_{\odot}$
(\cite{ferrarese02}).
Unfortunately, no other independent estimate of the
supermassive black hole mass exists for ESO509-IG066,
such as, for instance, those based on the [O{\sc iii}] line width or
on stellar velocity dispersion measurements. The
total AGN bolometric luminosity of the system is
$\sim$8.4$\times 10^{44}$~erg~s$^{-1}$, if a standard
ratio between the 1--10~keV and the bolometric luminosity
is applied
(\cite{elvis94}).
Even if one assumes the most conservative prescription for
the ratio between the galaxy circular and the halo
virial velocities (\cite{seljak02}),
the inferred mass of the latter is sufficient
to gravitationally bound the system if the active nuclei
are on the average accreting at
sub-Eddington rates (\cite{ferrarese02}).

X-ray observations allow
us to accurately probe the nuclear environment. With
the caveat of the small sample, it is
intriguing that all members of X-ray
detected AGN pairs suffer some degrees of absorption,
which in at least 50\% of cases might be Compton-thick
(\cite{komossa03,ballo04}). This may 
indicate that galaxy encounters are indeed effective
in driving gas to the nuclear environment.
If this is generally true, a certain number of AGN pairs
might be missed by
optical surveys due to obscuration of
one of the members (or both), which leads
to a wrong classification. These pairs
could show up only in hard X-rays.
ESO509-IG066 could be the prototype of a larger hidden
population of binary QSOs, still to be discovered by
deep high-resolution X-ray surveys.

\begin{acknowledgements}
This paper is based on observations obtained with XMM-Newton, an ESA
science mission with instruments and contributions directly funded by
ESA Member States and the USA (NASA).
This research has made use of the NASA/IPAC Extragalactic Database (NED) which is operated
by the Jet Propulsion Laboratory, California Institute of Technology,
under contract with the National Aeronautics and Space Administration.
Useful comments by the referee,
Dr.~D.Grupe, are gratefully acknowledged.
\end{acknowledgements}

\end{document}